# Screening lengths and osmotic compressibility of flexible polyelectrolytes in excess salt solutions

Carlos G. Lopez,[1, a)] Ferenc Horkay,[2, b)] Matan Mussel,[2)] Ronald Jones,[3)] and Walter Richtering[1)]

[1)]*Institute of Physical Chemistry, RWTH Aachen University, Landoltweg 2, 52056 Aachen, Germany*
[2)] *Section on Quantitative Imaging and Tissue Sciences, Eunice Kennedy Shriver National Institute of Child Health and Human Development, National Institutes of Health, 49 Convent Drive, Bethesda, MD 20892, USA*
[3)]*Material Measurement Laboratory, National Institute of Standards and Technology, Gaithersburg, MD 20899, USA*
a) lopez@pc.rwth-aachen.de
b) horkayf@mail.nih.gov

We report results of small angle neutron scattering measurements made on sodium polystyrene sulfonate in aqueous salt solutions. The correlation length ($\xi$) and osmotic compressibility are measured as a function of polymer ($c$) and added salt ($c_S$) concentrations, and the results are compared with scaling predictions and the random-phase approximation (RPA). In Dobrynin et al's scaling model the osmotic pressure consists of a counter-ion contribution and a polymer contribution. The polymer contribution is found to be two orders of magnitude smaller than expected from the scaling model, in agreement with earlier observations made on neutral polymers in good solvent condition. RPA allows the determination of single-chain dimensions in semidilute solutions at high polymer and added salt concentrations, but fails for $c_S \leq 2M$. The $\chi$ parameter can be modelled as the sum of an intrinsic contribution ($\chi_0$) and an electrostatic term: $\chi \sim \chi_0 + K'/\sqrt{c_S}$, where $\chi_0 > 0.5$ is consistent with the hydrophobic nature of the backbone of NaPSS. The dependence of $\chi_{elec} \sim 1/\sqrt{c_S}$ disagrees with the random-phase approximation ($\chi_{elec} \sim 1/c_S$), but agrees with the light scattering results in dilute solution and Dobrynin et al's scaling treatment of electrostatic excluded volume.

## I. INTRODUCTION

The effect of counterions on the conformation and structure of polyelectrolytes in solution has attracted a great deal of interest because of its importance in many biological processes such as DNA condensation[1], protein folding[2] or joint lubrication[3-4]. An understanding of the mechanisms that determine the conformation of charged macromolecules is required to tailor the properties of novel functional materials, including pharmaceutical[5-6] and food products[7-8]. It has been found that changes induced by monovalent salts can be described by the Poisson-Boltzmann model despite its serious limitations arising from the mean-field nature of the model.[9-13] For example, the Boltzmann distribution does not account for the finite size of ions and the size dependent ion-ion correlations and fluctuation contribution to the ion distributions. In certain systems, specific interactions between the polymer chains and the ions may also play a significant role, which introduces further complexities that make the interpretation of the experimental results more difficult.[14-18]

In semidilute polymer solutions, i.e. above the overlap concentration, polymer chains interpenetrate forming networks with a concentration dependent mesh size $\xi$, known as the correlation length[19]. The correlation length governs the thermodynamic, conformational and hydrodynamic properties of polymer solutions and gels.[19–32] The concentration and solvent quality dependence of the correlation length in semidilute and concentrated solutions of neutral polymers has been extensively studied, and interpreted by mean-field and scaling theories[19,21-23]. Polyelectrolytes in salt-free solution display markedly different behavior from those of neutral polymers. Owing to strong electrostatic repulsion along the backbone, they adopt highly extended conformations and their scattering function displays a correlation peak. In excess salt, electrostatic forces become short-ranged and their effect is expected to be similar to that of excluded volume in neutral polymers. A transition from highly extended rod-like conformation in salt-free solutions to expanded coils in excess salt has been observed for various systems in dilute solution[33-36]. Studies on the scattering properties of semidilute polyelectrolyte solutions with excess added salt are sparse, with most earlier literature focusing either on single chain properties[37-38] on the influence of specifically interacting multivalent counterions[39-41].

In the present work, the small angle neutron scattering (SANS) response of sodium polystyrene sulfonate (NaPSS) solutions containing large excess of monovalent counterions has been systematically investigated in the semidilute concentration regime. NaPSS is a well-suited model polymer to investigate ion-polymer interactions, particularly the effect of sodium counterions on the electrostatic interactions and molecular conformation, because no specific interactions between NaPSS and sodium ions have been reported in aqueous solutions.

We report experimental results for the correlation length and osmotic compressibility of NaPSS solutions as a function of the added salt concentration, primarily focusing on the excess added salt-regime. It is found that the results disagree with the predictions of the scaling theory.

The paper is organized as follows. The theoretical section is followed by a brief description of the materials and methods. Then we present the results of SANS measurements analyzed in terms of the Ornstein-Zernike equation. The influence of the polymer concentration and monovalent ion (sodium chloride) concentration is studied in semidilute NaPSS solutions.

## II. THEORY

Using simple scaling arguments, de Gennes derived the concentration dependence of $\xi$ of polymer solutions[20]:

$$\xi \approx R(c^*)\left(\frac{c}{c^*}\right)^\gamma \approx \begin{cases} Ac^{-0.77} & good\ solvent \\ Ac^{-1} & \theta-solvent \end{cases} \quad (1)$$

where $c$ is the polymer concentration (number of repeating units per unit volume), $R$ is the end-to-end distance of the polymer chain, $c^*$ is the overlap concentration, and $A$ is a constant depending on the monomer size, Kuhn length and thermal blob size. Scaling theory predicts that correlation blobs repel each other with an energy $kT$, where $k$ is the Boltzmann constant and $T$ is the absolute temperature. The polymer contribution to the osmotic pressure ($\Pi_p$) is:

$$\Pi_p \approx kT\xi_\Pi^{-3} \quad (2)$$

which describes the concentration dependence of $\Pi$ of neutral polymers both in good and $\theta$ solvents. [We use the subscript $\Pi$ to distinguish the correlation length obtained from the osmotic pressure from that determined from scattering measurements $\xi_{OZ}$, (see Eq. 9).] For neutral polymers, $\xi_\Pi$ is proportional and larger than $\xi_{OZ}$, with the proportionality constant increasing with increasing solvent quality.[20,21]

The correlation length of polyelectrolyte solutions depends on the polymer and added salt concentrations. Dobrynin's scaling model predicts[42]:

$$\xi(c_S) = \xi(0)[1 + 2c_S/(fc)]^{1/4} \quad (3)$$

where $\xi(0) \equiv \xi_{SF} = (b')^{3/2}c^{-1/2}$ is the correlation length in salt-free solution, $b'$ is the effective monomer length, $c_S$ is the added salt concentration and $f$ is the degree of dissociated counterions. For $c_S \gg fc$, Eq. 3 has the form $\xi \propto c_s^{1/4}c^{-3/4}$.

The osmotic pressure of polyelectrolyte solutions contains a contribution analogous to that of neutral polymers in good solvent ($kT\xi^{-3}$) and a term arising from counterion osmotic pressure, which may be approximated as:

$$\Pi_i \approx kT \frac{c^2}{4c_S/f^2 + c/f}$$

The osmotic compressibility arising from the polymer and counterions can, therefore, be given as:

$$\frac{d\Pi_p}{dc} = kT \frac{3\gamma}{A_\Pi^3} c^{-(3\gamma+1)} \qquad (4a)$$

$$\frac{d\Pi_i}{dc} = kT \frac{f^2 c}{2c_S} \qquad (4b)$$

where $\xi_\Pi = A_\Pi c^\gamma$ and $4c_S/f \gg c$.[43]

The structure factor at zero scattering wave-vector $q$ is:

$$S(0) = kT \frac{d\phi}{d\Pi} \qquad (5)$$

where $\phi$ is the volume fraction and $d\phi/d\Pi$ is the inverse osmotic compressibility. For neutral polymers in good solvent (excluded volume exponent $\nu = 0.59$) scaling theory predicts $\xi \propto c^{-0.77}$, $\Pi \propto c^{2.3}$ and $S(0)/c \propto c^{-0.31}$, which matches the experimental data.[19,22] However, the correlation length obtained from scattering and osmotic pressure measurements differs by a factor of $\approx 4$, which is not anticipated from the scaling theory. At high polymer concentrations, $\xi$ decreases to values well below the Kuhn length ($l_K$) of neutral polymers, which is incompatible with the scaling interpretation of the correlation length.[22,44]

In many experiments made on neutral polymers the structure factor of the polymer solution has been modeled by the random phase approximation (RPA) expression:

$$\frac{1}{S(q)v_s} = \frac{1}{\varphi N v_p P(q)} + \frac{1}{(1-\varphi)v_s} - \frac{2\chi}{v_s} \qquad (6)$$

where $v_s$ and $v_p$ are the volumes of the solvent and monomer molecules, respectively, $\phi$ is the volume fraction of the polymer and $P(q)$ is the polymer form factor, normalized to $P(0) = 1$, and $\chi$ provides a measure of the polymer-solvent interaction. For $1/R_g < q < 1/\xi$, the form factor can be approximated as $P(q) \approx 2/(qR_g)^2$, where $R_g$ is the radius of gyration. Equation 6 then gives:

$$\frac{1}{S(q)v_s} = \left[\frac{R_g^2}{2\varphi N v_p}\right] q^2 + \frac{1}{(1-\varphi)v_s} - \frac{2\chi}{v_s} \qquad (7)$$

In the range $1/\xi < q < 1/l_K$, the form factor depends on the solvent quality $P(q) \propto (qR_g)^{1/\nu}$, and Eq. 7 takes a similar form but with the term in square brackets being multiplied by $q^{1.7}$ instead of $q^2$.

In the present paper we interpret the polyelectrolyte solution data by eq. 6 understanding that this equation only appropriate for high salt concentrations, where polyelectrolyte chain aggregation is suppressed. For polyelectrolytes in excess salt solution, the $\chi$ parameter contains an intrinsic contribution ($\chi_0$), which reflects the polymer solvent interactions in the absence of Coulombic forces, and an electrostatic contribution, which we will consider in more detail in the Discussion section.

III. EXPERIMENTAL

Materials: Sodium styrene sulfonate (NaSS), sodium chloride and $D_2O$ were purchased from Sigma-Aldrich. Potassium persulfate (KPS) was purchased from VWR.[46] De-ionised (DI) water was obtained from a milli-Q source. Dialysis membranes were purchased from Spectra-Por. (The identification of commercial products does not imply endorsement by the National Institute of Standards and Technology nor does it imply that these are the best for the purpose.)

Synthesis: NaPSS was synthesized by free-radical polymerization of (NaSS) in aqueous media using KPS as an initiator. 70 mL of water was de-gassed for one hour and heated to 50 °C in a round bottomed flask. An aqueous solution of KPS was added and stirring was continued until it was fully mixed. The reaction was allowed to proceed for five hours, with nitrogen being bubbled continuously. The solution was cooled down and the polymer was precipitated by addition of fourfold methanol and excess NaCl. The polymer was then washed in methanol and re-dissolved in water. Solutions were extensively dialyzed against DI water to remove any residual salt and then freeze dried. Solutions from the dried polymer were prepared gravimetrically by assuming a polymer density of 1.65 g/mL.

Small Angle Neutron Scattering: SANS experiments were performed at the NGB 30m and NGB 10m Small Angle Neutron Scattering instruments at the NIST Center for Neutron Research (Gaithersburg, MD, USA). We employed sample-to-detector distances of 1.3 m, 4 m and 13.4 m with a wavelength of $\lambda = 6$ Å at the NGB 30m SANS instrument, which gave a $q$-range of 0.003 Å$^{-1}$ to 0.4 Å$^{-1}$. At the NGB 10m SANS instrument, sample-to-detector distances of 1.55 m and 5 m with a wavelength of $\lambda = 5$ Å were used, resulting in a $q$ range of 0.009 Å$^{-1}$ to 0.6 Å$^{-1}$. An empty cell reading was subtracted from the samples. Absolute calibration was made against a direct beam, according to NIST standard procedures. Samples were measured in Hellma cells of the QS series.

IV. RESULTS AND DATA ANALYSIS

Figure 1 shows the background subtracted SANS profiles of NaPSS solutions measured at different polymer concentrations at constant salt concentration (left figure) and at different salt concentrations at constant polymer concentration (right figure).

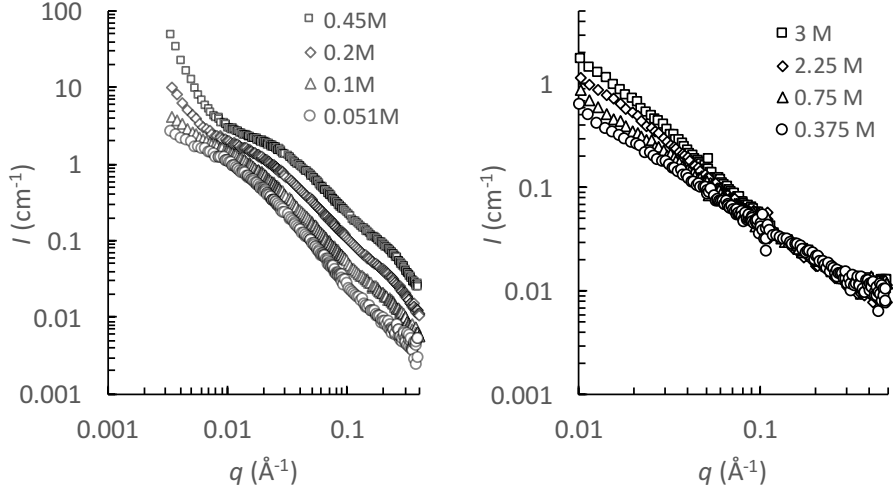

FIGURE. 1. Background subtracted scattering intensity for NaPSS in aqueous NaCl solutions. Left panel: $c_S = 3$ M, polymer concentrations are indicated in the figure. The lowest two concentration solutions were filtered (0.2 μm) prior to measurement. Measurements made by the NGB 30m instrument. Right panel: $c = 0.1$ M, salt concentrations are indicated in the figure. Measurements made by the NGB 10m instrument.

The scattering intensity relates to the structure factor as:
$$I(q) = \left[\frac{b_m}{v_m} - \frac{b_s}{v_s}\right]^2 S(q) \qquad (8)$$
where $b$ and $v$ are the scattering length and partial molar volume, respectively, and the subscripts m and s refer to the monomer and solvent. In Eq. 8 the scattering arising from salt is neglected.

In analogy to neutral polymers, the scattering of a polyelectrolyte solution in excess added salt can be described by a Lorentzian function:
$$1/I(q) = A + Bq^2 \qquad (9)$$

where $A$ and $B$ are fit parameters, related to the zero angle scattering intensity and correlation length as $I(0) = 1/A$ and $\xi = B/A$.

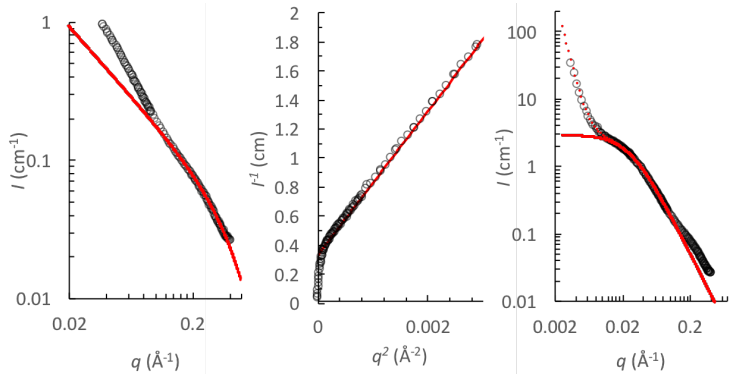

FIGURE. 2. Fitting procedure to determine the correlation length and zero-scattering intensity. Circles: coherent scattering intensity for sample with $c = 0.38$ M, $c_S = 3$ M. Left: Fit to worm-like

chain form factor at high $q$, red line is $P(q) = \pi/(b'q)e^{-q^2R_C^2/4}$. Middle: Fit to Lorentzian function (first term of Eq. 9). Right: Fit to Eq. 9 with clustering term (dotted line) and to Lorentzian term only (full line).

The fitting procedure to estimate the background (including incoherent and other $q$-independent scattering contributions), correlation length and clustering term is illustrated in Figure 2. The high $q$ region is fitted to a worm-like chain form factor in the $qL_K > 1$ limit: $S(q) \approx P(q) = \pi/(b'q)e^{-q^2R_C^2/4} + Bkgd$, where $b'$ is the effective monomer length, $R_C$ is the cross-sectional radius of the chain (set to 0.4 nm[46]) and $Bkgd$ is a constant that accounts for the $q$-independent scattering. A representative fit to this equation is shown in Fig. 2 (left figure). The correlation length is estimated by fitting the Ornstein-Zernike function (Eq. 9) in the mid-$q$ region, as illustrated in Fig. 2b. Finally, a power-law term $I(q) = Dq^{-m}$, where $D$ and $m$ are constants, is added to account for the excess scattering at low-$q$.

## V. DISCUSSION

### A. Scaling Analysis

Figure 3 shows the correlation length and reduced zero angle scattering intensity as a function of polymer concentration in solutions of different salt contents. Both quantities decrease as power-laws of the polymer concentration at constant salt concentration, shown by the best-fit lines. Scaling model predicts that $\xi$ and $I(0)$ are related through Eqs. 2, 4 and 5. Applying these equations, we find that the calculated values of $I(0)$ are nearly two orders of magnitude smaller than the measured ones. This discrepancy indicates that $\xi_{OZ} < \xi_\Pi$, which has been reported for several neutral polymer systems[22,23].

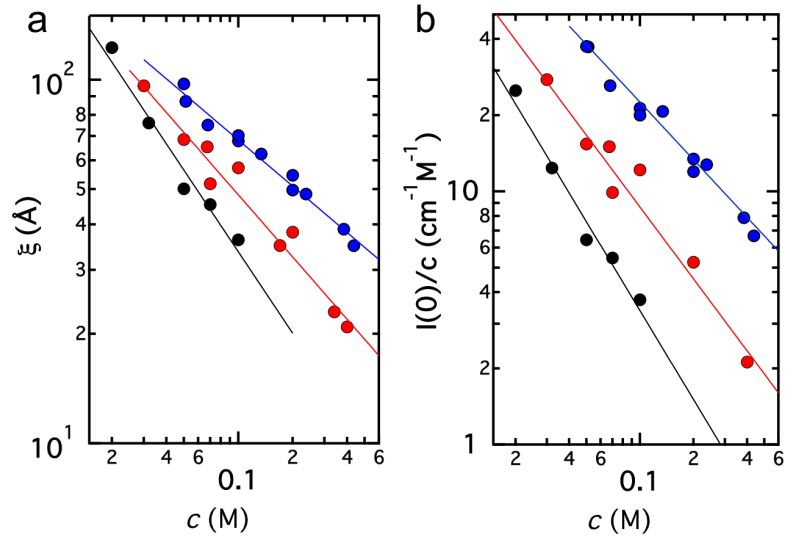

FIGURE. 3. Correlation length (a) and reduced zero angle scattering intensity (b) as a function of the polymer concentration at different concentrations of added salt. Symbols: 3 M (blue), 1.5 M (red) and 0.375 M (black).

In Fig. 4a the exponent $\gamma$ obtained from the concentration dependence of $\xi$ ($\propto c^{\gamma}$) is plotted as a function of $c_S$. For $c_S = 0.375$ M, the data display the expected[42] scaling dependence of $\xi \propto c^{-0.75}$. For higher $c_S$, the concentration dependence becomes weaker, in agreement with experimental results obtained for flexible neutral polymers in good solvents such as polystyrene in dichloromethane or PDMS in toluene[21,22]. As $c_S$ increases, the solutions approach the θ point ($\simeq 4.17$ M NaCl at $T \simeq 290$ K[47]), and scaling theory predicts that $\gamma$ should decrease from -0.77 at low added excess salt concentration (i.e. $c_S \ll 4.17$ M, $c_S \gg fc/2$) to $\gamma = -1$ at the θ state. However, in the present system the opposite trend is observed: the exponent increases to a value of $\gamma \simeq -0.45$ at $c_S \simeq 1.5 - 3$ M. As discussed below, this feature can be accounted for by the RPA method.

Figure 4b plots the ratio $\xi_\Pi/\xi_{OZ}$ required to match the measured and calculated osmotic compressibilities as a function of the added salt concentration. At high $c_S$, $\xi_\Pi/\xi_{OZ} \simeq 4.2$ is found, which is close to the value 3.8 reported for polymers in good solvent[22,23]. Using these ratios, the observed and calculated exponents for the variation of $I(0)$ with the polymer concentration are in reasonably good agreement (see Fig. 4c).

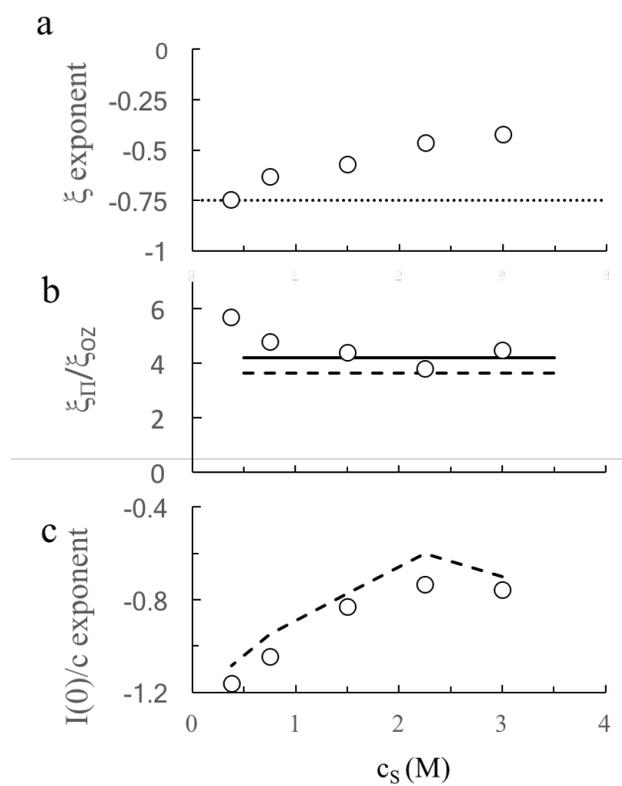

FIGURE. 4. a) Exponent for the $c$ dependence of $\xi$, dotted line is scaling prediction for neutral polymers in good solvent and polyelectrolytes in the presence of excess added salt. b) Ratio of scattering ($\xi_{OZ}$) and osmotic ($\xi_\Pi$) correlation lengths. Full line: average value at high salt concentration, dashed line: value for neutral polymers in good solvent[23], c) Exponent for the $c$ dependence of $I(0)/c$. Data points are experimental values, dashed line shows the variation of the exponent calculated from Eqs. 2, 4 and 5.

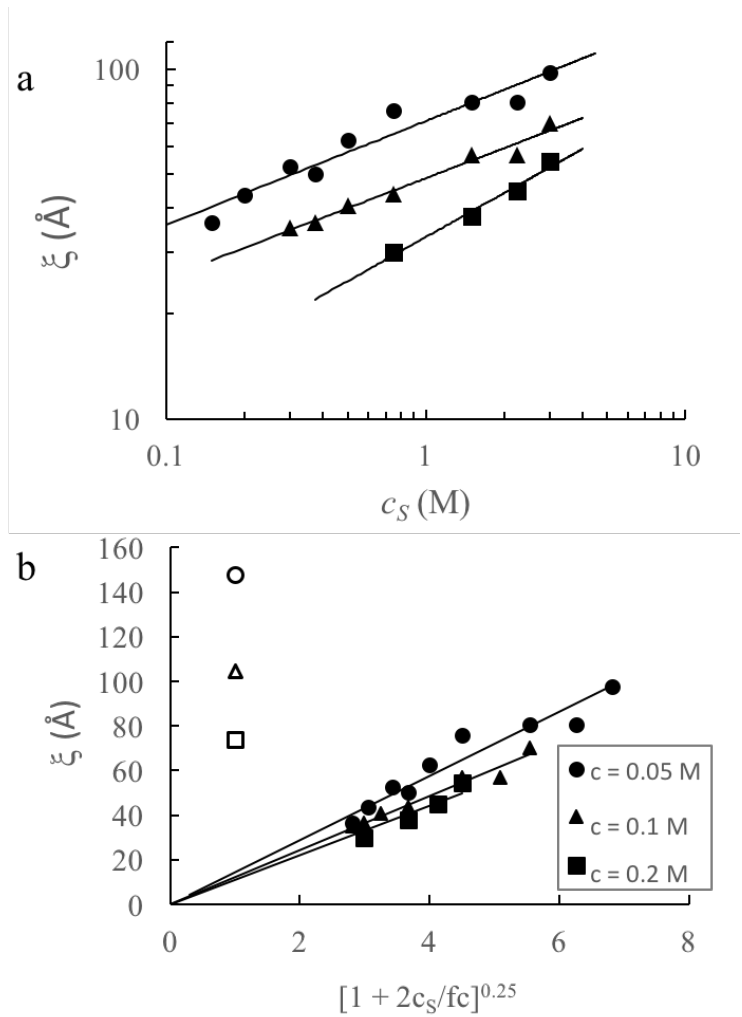

FIGURE. 5. a. Dependence of $\xi$ on the added salt concentration for different polymer concentrations. b: Scaling plot according to Eq. 3. Filled symbols are data measured in excess salt, hollow symbols are for salt-free solutions, calculated from $\xi = 33c^{-1/2}$ Å.[48,49] Symbols have the same meaning in parts *a* and *b*.

The salt-dependence of the correlation length at constant polymer concentration is shown in Fig. 5a. The best fit exponents for $c = 0.05, 0.1$ and $0.2$ M, are $-0.30 \pm 0.06$, $-0.28 \pm 0.04$ and $-0.42 \pm 0.13$, respectively. These values exceed the scaling prediction of 0.25 (see Eq. 3). In figure 5b $\xi$ is plotted as a function of $[1 + 2c_S/fc]^{1/4}$. According to Eq. 3 this plot should be a straight line going through the origin, which is approximately observed for data measured in excess salt. The values of $\xi$ in salt-free solution, calculated as $\xi_{SF} = 33c^{-1/2}$,[48,49,50] are 5 – 7 times larger than the values extrapolated from the excess salt data. Scaling defines the correlation length as being equal to the end-to-end distance of a chain at the overlap concentration, i.e. $\xi(c^*) = 6^{1/2}R_g(c^*)$ for Gaussian chains. On the other hand, comparison of Eq. 9 with the Zimm approximation yields

$\xi(c^*) \simeq R_g/3$. The correlation lengths obtained from Eq. 9 are therefore expected to be much smaller than the $\xi$ calculated by scaling. Adjusting the results of Eq. 9 by a factor of $3 \times 6^{1/2} \approx 7.3$ would bring the excess-salt and salt-free data plotted in Fig 5b to agreement.

### B. RPA and Double screening

Given the discrepancies between experimental data and scaling theory, in particular with respect to the dependence of the correlation length on the polymer concentration, we compare our experiments with the random phase approximation method, which is known to provide a reasonably good description of scattering from concentrated neutral flexible polymer solutions.

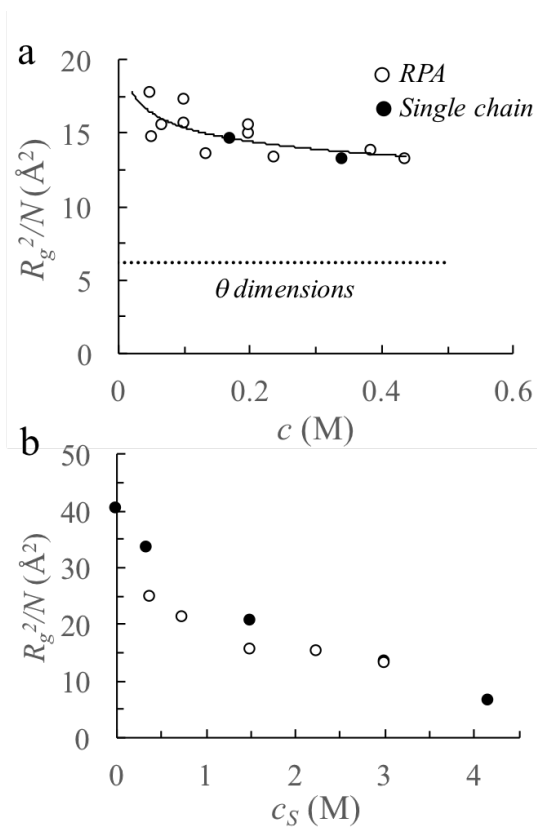

FIGURE. 6. Chain dimensions calculated from Eq. 7 (hollow symbols) and from single chain measurements (filled symbols) as a function of polymer concentration in 3 M NaCl (a) and as a function of added salt concentration for $c = 0.34$ M (b). The datum at $c_S = 4.17$ M corresponds to NaPSS at the $\theta$ condition 17.8 °C,[47], also indicated in part a) by the dotted line.

Equation 7 predicts $\xi \sim c^{-1/2}$ and $I(0)/c \sim c^{-1}$ in excess salt if $R_g^2/N$ and $1/(1-\phi) - 2\chi$ do not vary strongly with the polymer concentration. These dependences are consistent with the experimental results obtained for NaPSS in 3 M NaCl. Figures 6a and 6b compare the values of $R_g^2/N$ calculated from Eq. 7 with direct measurements by Spiteri[51] using the zero-average contrast (ZAC) method as a function of polymer and added salt concentrations, respectively. For $c_S = 3$

M, good agreement is found between the two methods. However, for lower added salt concentrations, the RPA consistently underestimates the dimensions of NaPSS chains. For all concentrations studied here, the polymer chains are swollen with respect to the θ dimensions reported by Hirose et al[47] at $c_S$ = 4.17 M, at $T$ = 17.4 °C. It is unclear whether the larger chain dimensions at $c_S$ = 3 M are primarily the consequence of excluded volume effects or the increased persistence length of the chains.

Muthukumar's double screening theory[52] predicts $\xi \sim c^{-1/2}$ at low added salt concentration, $\xi \sim c^{-3/4}$ at high salt and moderate polymer concentrations, and $\xi \sim c^{-1/2}$ for concentrated polyelectrolyte solutions in excess salt. These predictions are consistent with the experimental results reported in the present work. The zero-angle scattering intensity may be expressed in terms of the $\chi$ parameter using Eq. 6. The RPA method and Muthukumar's double screening theory predict:

$$\chi = \chi_0 - \frac{K}{c_S} \qquad (10)$$

where $K$ is a constant, that depends on the relative permittivity of the solvent and the Kuhn length of the polymer.[45,52]

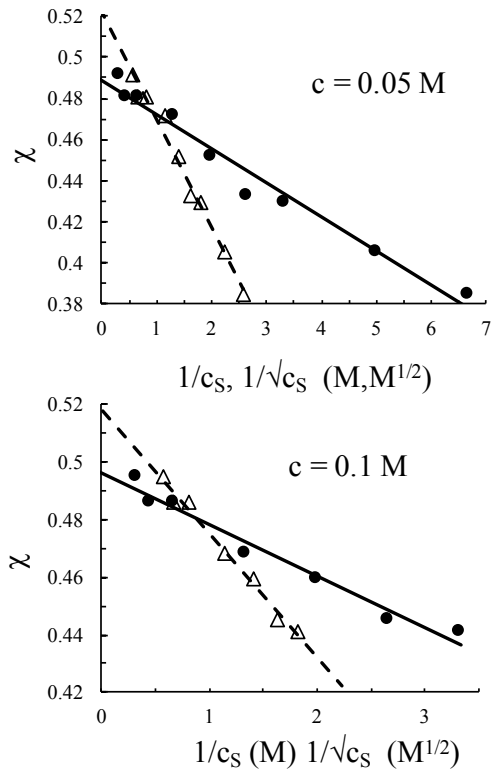

FIGURE. 7. The $\chi$ parameter as a function of $c_S$ (full circles) and $\sqrt{c_S}$ (open triangles) for $c$ = 0.05 M (top) and $c$ = 0.1 M (bottom). Lines are best linear fits to the data.

In Figure 7 the $\chi$ parameter is plotted as a function of $1/c_S$ for $c$ = 0.05 and 0.1 M. The linear relation predicted by Eq. 10 is observed for all polymer concentrations studied. The $1/c_S = 0$ intercept yields $\chi_0 \simeq 0.49$, which is too low given the hydrophobic nature of the polystyrene backbone. Assuming instead a dependence of $\chi = \chi_0 - K'/c_S^{1/2}$ gives a more reasonable estimate $\chi_0 \simeq 0.52$. Parameters $K$ and $K'$ are found to be 0.017 and 0.05, respectively.

An earlier estimate of $\chi_0 \approx 1.1$ was obtained by Prabhu et al[40] by extrapolating $\chi$ vs. $1/c_S$ to infinite ionic strength for NaPSS in excess barium chloride. The larger value of $\chi_0$ can be explained as arising from two factors: First, the NaPSS of reference 40 was synthesized by sulfonation of polystyrene, giving a degree of sulfonation of 96%, instead of 100% obtained for samples made by radical polymerization of styrene sulfonate[47], as is the case in the present work. The small fraction of non-sulfonated polystyrene is known to lead to greater backbone hydrophobicity, which manifests itself in, for example, smaller chain dimensions in dilute excess-salt solution and a lower $\theta$-salt concentration[47,53]. A second factor that may account for the discrepancy is that $Ba^{2+}$ cations interact specifically with the sulfonate groups, leading to different chain conformations and phase behaviour[40,41,54,55]. It is possible that $\chi$ exhibits a different $c_S$ scaling in the presence of NaCl and $BaCl_2$, thus complicating a comparison of the extrapolations to infinite ionic strength.

The dependence of the $\chi$ parameter on the square root of the added salt concentration is consistent with results reported for various polyelectrolyte systems[47,56-60], and is also consistent with Dobrynin et al's[42] and Odijk et al's[61] treatment of excluded volume in excess salt solutions, both of which expect the excluded volume strength to vary linearly with the Debye screening length.

## VI. CONCLUSIONS

We have evaluated the correlation length and osmotic compressibility of polyelectrolyte solutions in excess salt. Scaling theory correctly describes the variation of the correlation length with the polymer concentration at low and moderate added salt concentrations, but at high salt concentrations the RPA and double screening theory work better. The ratio of the scattering correlation length and the osmotic correlation length is found to be $\simeq 4$ at high added salt concentration, in agreement with earlier reports for neutral polymers in good solvents. The correlation lengths obtained from the peak position in the scattering profiles of salt-free polyelectrolytes and from the fit to a Lorentzian function differ by an order of magnitude. The electrostatic $\chi$ parameter is found to vary as $\chi_{elec} \sim 1/c_S^{1/2}$, in contrast with the linear dependence predicted by various theories.

**Acknowledgements**

F.H and M.M. acknowledge the support of the Intramural Research Program of the NIH, NICHD. Access to the NGB 30m SANS was provided by the Center for High Resolution Neutron Scattering, a partnership between the National Institute of Standards and Technology and the National Science Foundation under Agreement No. DMR-1508249. Use of the NGB 10m SANS was supported by the NIST nSoft Consortium. We thank Dr. B. Hammouda for his excellent advice.